# Towards a Conceptual Approach of Analytical Engineering for Big Data

Rogério Rossi and Kechi Hirama

*Abstract*—Analytics corresponds to a relevant and challenging phase of Big Data. The generation of knowledge from extensive data sets (petabyte era) of varying types, occurring at a speed able to serve decision makers, is practiced using multiple areas of knowledge, such as computing, statistics, data mining, among others. In the Big Data domain, Analytics is also considered as a process capable of adding value to the organizations. Besides the demonstration of value, Analytics should also consider operational tools and models to support decision making. To adding value, Analytics is also presented as part of some Big Data value chains, such the 'Information Value Chain' presented by NIST among others, which are detailed in this article. As well, some maturity models are presented, since they represent important structures to favor continuous implementation of Analytics for Big Data, using specific technologies, techniques and methods.  Hence, through an in-depth research, using specific literature references and use cases, we seeks to outline an approach to determine the Analytical Engineering for Big Data Analytics considering four pillars: Data, Models, Tools and People; and three process groups: Acquisition, Retention and Revision; in order to make feasible and to define an organization, possibly designated as an Analytics Organization, responsible for generating knowledge from the data in the field of Big Data Analytic.

## I. INTRODUCTION

Data avalanche is one of the strongest features for a paradigm that is currently identified as Big Data and collaborates on its conceptual and practical approaches. Data are essential for determining Big Data paradigm and for generating value for organizations [1], but the Big Data paradigm is also determined by the variety of structured and non-structured data and by the data generated by many types of data-based systems, such as transactional systems, Web Systems, IoT systems, etc.

Big Data can be considered in a variety of sectors, such as business, industry and commerce, education and healthcare. Transport, logistics, and finance, as well as other sectors, also consider this recent paradigm of the information era [2]. Scientific programs, some with greater and others with less emphasis, have also benefited from the technical and technological characteristics related to Big Data.

Big Data, as a recent paradigm of the information era, presents a relevant phenomenon related to the Datafication Phenomenon [3], which determines that words, physical locations and human interactions have become data. This phenomenon highlights the importance of the data generated by the various data systems and can be used and reused by the society to generate results that favor decision-making [4].

To manage this vast volume of data, organizations have adopted technologies, tools and models associated with Big Data and, therefore, also adopt another relevant capability that refers to data analysis, or Analytics, sometimes also referred to as Big Data Analytics [5].

Analytics presents some relevant conceptual features, but important practices regarding its capacity for the Big Data domain can be verified in [6] and [7]. Analytics is a simplified form of referring to Big Data Analytics and corresponds to data analysis activities for Big Data [8] and [9]. In the scope of this research, the concept of Analytics must be associated with the Big Data paradigm, seeking to favor a vision for implementing Big Data Analytics in an organization that aims to be data-driven.

Data presented in large volumes, with varied types (structured, unstructured, semi-structured) being generated and providing high-speed responses to favor decision-making, need to be analyzed. This analysis is a broad question within the Big Data domain, perhaps of such breadth and magnitude that it can also support the word 'big' from the term Big Data, which is sometimes observed exclusively by the vast volume of data.

Given the capacity and strength that analytical activity has for the Big Data domain, its conceptual and practical approach, with new techniques, technologies, tools and models, enable what can be considered the semantic part of Big Data [10], that is, the results and the responses that these data are capable to offer [11].

However, there are critical issues to consider for Big Data [12], which require that data should be analyzed by organizations that already have the Big Data paradigm. Regardless of the size or the sector to which they pertain, organizations may present some difficulties in implementing and using analytics in the Big Data domain. A specific organization can favor and meet the needs that companies seek for strengthening data analysis in Big Data. However, implementing this specific organization requires investments and efforts to favor data-oriented results, being considered of high relevance and impact for companies of many sectors [13].

A vital way to add value to results, data must be properly managed: from their capture, considering their storage, validation and culminating in their analysis and availability for visualization. This kind of data cycle has favored Big Data Value Chain proposals, such as those in [1] and [14].

Manuscript received November 13, 2019; revised February 12, 2020. This work was supported by Programa de Educação Continuada em Engenharia (PECE) (Engineering Continuing Educational Program) of Polytechnic School of University of São Paulo.

R. Rossi and K. Hirama are with the Department of Computer and Digital System Engineering, Polytechnic School of University of São Paulo, SP Brazil (e-mail: rogeriorossi8@gmail.com, kechi.hirama@usp.br).

The analytical approach for Big Data can also be institutionalized with the support of some maturity models that seek to provide capacity for data management. Thus, some maturity models for Big Data Analytics, such as those in [15], [16] and [17] are presented, aiming at boosting the generation of value in data-driven organizations.

Based on value generation and the need of structuring data analysis activities in an organization to manage Analytics for Big Data, the scope of this research is highlighted, and seeks to integrate the principles of engineering to the activities of data analysis and data management. Engineering is a dynamic discipline capable of adapting the society needs in the most varied areas, such as aerospace, chemistry, civil, etc. [18], and it can structure and enable relevant activities to manage and to analyze data in a data-driven organization.

In general, engineering fundamentals are able to provide means for solving open problems, that is, problems that do not have a single solution, and can also be effective in data analysis processes. Data analysis in the Big Data domain is an open problem that can be modeled to provide a suitable solution, but not by a single path. The scientific basis for Big Data Analytics relies heavily on mathematical models, statistical principles, and complex computational models that allow constructing solutions that enable data analysis and their consequent use.

Thus, the main objective of this research is presenting a conceptual approach to Analytical Engineering for Big Data based on scientific knowledge, definitions and concepts of the literature, and through use cases. This research seeks to determine a structure to integrate the four pillars defined to Analytical Engineering for Big Data that correspond to Data, Models, Tools and People and its three process groups corresponding to Acquisition, Retention and Revision.

This conceptual approach to Analytical Engineering, considering its four pillars, can contribute to the deployment of the expected results of a data-driven organization responsible for managing data to cover their capture, the analysis itself, as well as the availability for the results presentation, which can be called Analytics Organization.

Providing a conceptual approach of Analytical Engineering for Big Data has enabled us to define the activities of the research, reported in the following sections: (two) conceptually approaching Analytics for Big Data and the value that the organization can add by using value chain structures and other types of specific structures for Big Data Analytics; (three) presenting analytics for Big Data domain, specifically their categories and considerations that integrate Big Data and Analytics; (four) presenting the structure that determines the conceptual approach of Analytical Engineering for Big Data, detailing each of the pillars and the process groups that determine its structure; and, finally, (five) presenting this research conclusions, as well as suggestions for future work

## II. ANALYTICS VALUE

Data analysis for Big Data or Analytics for short, as it is commonly referred in a simplified and emphatic way for the Big Data domain, is a process capable of adding value to the organization in which data analysis is performed to favor decision-making.

For [19], although a program to implement analytics can provide several benefits to the organization, it should not be started without a strategic direction. For [20], an organization must make two considerations for adopting analytics for Big Data: 1) demonstration of value, and 2) operationalization. It is therefore necessary to determine the viability and value-added of the business and the due commitment of the sponsors, as well as the disclosure of the benefits. They should also consider operational instruments that enable a proper cycle of development and delivery, to favor its use.

For [21], analytics is a process that adds organizational value or means of generating value and return in various sectors, such as healthcare, manufacturing, public sector, among others. Considered a set of specific activities, technologies and techniques, Analytics is present in many Big Data value chain structures, and is generally considered one of its phases.

In general, value chains support the strategic level of an organization and guide business from the generation of value. The concept proposed by [22] for using the value chain states that it considers a series of activities to create and to construct value. Applied to the Analytics for Big Data, the value can be driven indirectly because the analysis results are completed by a decision maker who needs an external capability (data analysis) to support decision making.

Therefore, some proposals for value chain structures for Big Data domain have been presented to provide means to verify the value added through the data analyzed. [1] present the 'Data Value Chain'; [23] presents the 'Information Value Chain'; [14] presents the 'Big Data Value Chain'. These value chain structures present some common phases and essentially culminate in the data analysis phase, that is, it culminates in the Analytics Phase, as a means of presenting the results of the analyses made for later availability and visualization.

Considering the aforementioned Big Data value chain structures, the 'Data Value Chain' presented by [1] treat Analytics as part of the 'Data Exploitation' phase, considered the heart of the value chain and also one of the phases with more mature tools and techniques. [14] proposes in the 'Big Data Value Chain' a phase named Analytics and which encompasses recent technologies and techniques that favor data analysis for Big Data. Web Mining, Text Mining, and Statistical Analysis are examples of techniques considered at this stage of the 'Big Data Value Chain' proposed by [14]. The authors define 'data analysis' as the use of tools and analytics methods to inspect, transform, and model the data to extract value. [23], in its proposal called 'Information Value Chain', also suggests a specific phase for analytics in which it considers a framework that predicts infrastructure, platform, and specific processing mechanisms to support data analysis actions.

These big data value chain proposals come in favor of validating the value of all the components related to Big Data Systems [24], including Analytics, which is a way of generating value to the organization through the results presented after the analysis of massive datasets with varied data types.

Other ways of approaching Big Data Analytics can be verified, according to [6], [11], [15], [16], [19], [25], [26] and, which present some structures related to Big Data Analytics specifically, or sometimes, structures that are more comprehensive and directed to Big Data Systems. Some of these proposals correspond to maturity models for Big Data and present relevant aspects of data analysis, such as the models presented by [15] and [16].

In general, maturity models that have been widely presented for the Information Technology area converge to the technological, organizational or process-oriented aspects. For the Big Data domain, maturity models are represented by maturity levels and propose a continuous improvement of the processes related to the Big Data and Analytics domains, denoting a proposal for continuous process improvement.

Reference [16] presents the 'Big Data Maturity Model', which considers five maturity levels: 1) Catching Up, 2) First Pilot(s), 3) Tactical Value, 4) Strategic Leverage, and 5) Optimize & Extend. At the 'Catching Up' maturity level, trainings and understandings about Big Data, begin to move towards the first pilots to be treated at the 'First Pilot' maturity level. At level three, 'Tactical Value', the actions of analytics are emphasized in some projects and it is possible to present some evidence of return on investment. For levels four and five, 'Strategic Leverage' and 'Optimize & Extend', respectively, the author proposes that Big Data can be treated as a capability for business strategy. The focus on completing the maturity levels of the proposed model provides optimization capability to the organization's business models.

With five maturity levels, the 'TDWI Big Data Maturity Model' presented by [15] addresses issues related to Data Management, Analytics and Governance, demonstrating that the strength of the proposed model lies in data management and in its analysis. 'TDWI Big Data Maturity Model' also considers five maturity levels: 1) Nascent, 2) Pre-adoption, 3) Early adoption, 4) Corporate adoption, and 5) Mature / Visionary. In its fifth level 'Mature/Visionary', the model defines that analytics must present an organizational maturity whereby analytics is integrated with the Business Process, yielding results based on all types of data, including real-time data, to support critical missions of business.

DWCMM (Data Warehouse Capability Maturity Model), proposed by [25] is another maturity model that considers technical aspects of the Data Warehouse and Business Intelligence solution and a set of organizational processes that sustain them. DWCMM features five maturity levels: 1) Initial, 2) Repeatable, 3) Defined, 4) Managed, and 5) Optimized. The six following categories are presented for each maturity level: 1) Architecture, 2) Data modeling, 3) ETL (Extract, Transform, Load), 4) Business Intelligence applications, 5) Development Process, and 6) Service process.

These examples of maturity models were preceded by a specific model for data quality management presented by [26] called 'Data Quality Management Maturity Model' aiming at enabling maturity to (structured) data management through its four maturity levels: 1) Initial (DBMS Management), 2) Defined (Data Model Management), 3) Managed (Standard Metadata Management), and 4) Optimized (Data Architecture Management).

References [6], [11] and [19], presented some other types of structures as detailed below. They do not correspond to maturity models or value chain structures, but are intended to support the actions of analytics for Big Data within the organization, providing conditions for self-evaluation for organizations that already act or intend to work with the practices, techniques and technologies associated with Big Data Analytics.

Reference [19] proposes a specific framework for organizations to perform a self-assessment in relation to aspects of Big Data, describing three capacity levels: 1) Aspirational, 2) Experienced, and 3) Transformed, highlighting that the 'Transformed' capacity level should be considered by organizations with experience in applying analytics to a diverse set of organizational functions.

The structure proposed by [6] also presents mechanisms to identify how an organization is in relation to the adoption of Big Data, presenting the following stages: 1) Educate, 2) Explore, 3) Engage, and 4) Execute. At the 'Execute' stage, the organization is considered to have already defined and delivered (deployed) two or more Big Data initiatives and continued to apply advanced analytics.

Reference [11] presented a structure with three levels, with greater emphasis on analytics maturity, which allow verifying: 1) Analytical Innovators, 2) Analytical Practitioners, and 3) Analytically Challenged. Each of these analytics levels represents innovative organizations that have already met some relevant challenges and having a data-driven organizational culture; for the first level, organizations that already use Analytics for some effective operational improvements; for the second level, organizations that have relevant challenges to be data-oriented; and, for the third level (intuition) for organizations that still have a managerial profile for intuition-based instead of data-based decision making.

Although value chain proposals and other types of structures have been presented for Big Data Analytics, an issue was presented by business executives seeking to add value to organizations from the data: "Is not 'Big Data' just another way of saying 'analytics'?" as mentioned by [27], which intensifies the relevance of the value that analytics determines for the Big Data scenario. For [27], Big Data Management is responsible for seeking to glean intelligence from data and for translating that into business advantage. This proposition retakes the ability of analytics, which corresponds to a group of technologies and techniques, to meet the needs of the business, generating intelligence to the business from data analysis.

Analytics as a means of generating value for the organization, as a transformer mechanism, receives differentiated conceptualizations, but the term does not present a consensual definition. In a comprehensive way, 'Analytics is the extraction of knowledge from information'; however, other concepts can be observed. For [14] data analysis comprises analytical methods and tools to transform and to model data to extract value. [28] consider that data analysis is the final phase of the Big Data value chain, in order to extract values, providing suggestions and values. For [23] analytics is verified to be the synthesis of knowledge from information.

Thus, the implementation of analytics as a transforming process in the organization, whereby data can be transformed into knowledge to favor decision makers, is recognized through the analysis of massive datasets of different data types, thus evidencing the generation of information and knowledge using data. Specific techniques and technologies are integrated to transform data into knowledge, determining the data mining paradigm [29]; with the application of complex and varied types of algorithms, data mining has been widely used by the information industry to foster the generation of knowledge from data.

The application of data mining techniques so as to enable intelligence for business determines the concept of Business Intelligence (BI), a term that became popular in the information industry in the 1990s. BI and Analytics presented relevant conceptual proximities and convergent practices; yet the information industry in the Big Data era determines the term Big Data Analytics, sometimes addressed exclusively as Analytics, to refer to the actions, models, technologies, techniques, and people employed in data mining tasks to provide the 'best' knowledge for business areas, thus determining the 'best' result for the business to favor decision making.

Thus, considering the capacity of the data as input for decision making, analytics corresponds to a set of practices capable of adding value to the organization. In this sense, the next section presents some categories defined for Analytics for Big Data.

### III. CATEGORIES OF ANALYTICS FOR BIG DATA

Big Data refers to a paradigm with some common properties identified by several authors, such as [6] and [14] and with some variations, as verified in [6] and [30]. These common and consensual features refer to its three main properties – volume, variety and velocity. The variations allow verifying other properties, as proposed by [14] and [30], who also consider the 'value' property, and [6] that consider the 'veracity' property.

Therefore, Big Data is characterized by specific properties considering a vast volume of data. It demands the analysis of these data, culminating in what is called Analytics, a term that has sometimes been mistaken by business executives with the term Big Data itself and which is generally treated as 'Big Data Analytics'.

The main properties and features intrinsic to Big Data relate this to an analytical need, in which data analysis offers primordial value to organizations. According to [31], analytical thinking can be considered one of the critical aspects of the Big Data paradigm, that is, thinking data-analytically is vital not only to the professionals involved with these types of analyses, but also for the whole organization.

Analytics is the extraction of knowledge from information, and, specifically for the Big Data domain, 'Big Data Analytics' describes analytical techniques applied to Big Data systems that are so large that they require advanced technologies for data storage, analysis and visualization [32]. For [33] 'Big Data Analytics' differs in data input, data access patterns and the parallelism required and online (streaming) applications incur significant latency costs.

For [34] analytics represents the overall process of extracting insights from Big Data. The processes for extracting knowledge from the data can be represented in five stages, divided into two main groups: 'Data Management Group' and 'Analytics Group'. For the 'Data Management Group' the processes are: 1) Acquisition and Recording, 2) Extraction, Cleaning, and Annotation, 3) Integration, Aggregation, and Representation; and for 'Analytics Group' the processes are: 4) Modeling and Analysis and 5) Interpretation.

However, for the Big Data domain, there are several categories of analytics that address the needs related to data analysis. As examples, text analyses are performed using techniques categorized in Text Analytics; data analysis activities provided by data from mobile devices and their various types of applications are executed by Mobile Analytics techniques; data analysis from different types of social networks occur by using Social Analytics techniques. Thus, it is possible to highlight some categories for analytics as specified by [14], [28] and [32].

The authors converge in most of the categories, which for [28] are defined as Big Data Analysis Fields; however, there are differences according to the proposals, especially when verifying the proposals in [32] in relation to the proposals in [14] and [28].

These categories favor the aggregation of technologies, techniques and models to be used for mining and any subsequent action performed in the data analysis and for visualization. In this sense, specific techniques and tools for data mining in the Big Data era (petabyte era) have challenges that focus on three classes presented by [35] the first being considerations on 'data processing and computing', the second related to 'data privacy and domain knowledge', and the third concerning 'Big Data mining algorithms'.

Data analysis considering the Big Data era requires revision in data mining properties, also considering the semi-structured and unstructured data. Hence, [36] present challenges and opportunities for what they call 'Big Data Mining' that must deal with heterogeneity, extreme scale, velocity, privacy, accuracy, trust, and interactiveness. [35] also mention that many data mining algorithms require that the entire dataset should be stored in the main memory to generate the due results, which is a classic barrier to data mining in the Big Data era.

Considering that data mining has to be reviewed for the Big Data, it can favor the convergence of models and technologies that consider these different categories of Analytics proposed by [14], [28] and [32], to determine the best results. This convergence should be made possible to solve problems involving many types and varieties of data, and, in many cases, the use of more than one analytics category, i.e. problems involving, for example, analysis of texts, structured data, and Web data will all be natural to the Big Data scenario (petabyte era).

Therefore, professional qualification is also necessary, with specific skills for doing these activities in the technological or in the business and application areas. The ability to capture, to store, and to manage data with specific technologies of varying expertise should enable business models along with data models to generate value from massive datasets. As mentioned by [33], different applications require different data-processing techniques and optimization, which are also related to the techniques and technologies that make analytics viable.

Thus, Analytical Engineering is presented in the next section to support an Analytics Organization and its main

functions, as well as to support activities and development of Analytics Deployments to solve data-driven business problems.

IV. ANALYTICAL ENGINEERING: A CONCEPTUAL APPROACH FOR BIG DATA

Reference [20] proposes that before defining its schema for Big Data Analytics, an organization should define an appropriate Big Data Strategic Plan. The same emphasis on Strategic Planning for Big Data Analytics is highlighted by [37], that says that the strategic planning should propose ways to define why, how, and when Big Data Analytics technologies are appropriate practices for the organization; managing business users' expectations and defining an organizational action plan for Big Data Analytics implementation.

Given that planned practices and actions enable the organization's strategy to become data-driven, specific engineering concepts can be applied to model problems and solutions by considering datasets, tools, specific models and people skills, in order to provide appropriate solutions.

For the Big Data domain, data should be analyzed considering at least three of its properties: volume, variety and velocity; possibly adding the value property. In this sense, problems and proposed solutions must consider these properties using data as input, turning them into results that enable the business solutions, that is, that allow solving organizational problems.

A diversity of problems in the contemporary world are modeled and proposed using engineering tasks. Humanity has witnessed significant changes throughout engineering [38]. Based on science and mathematics, engineering considers and defines mechanisms for many types of problems presented in several areas, such as industry, environment, finance, etc. The systematization of models and standards using diversified techniques and technologies allows transforming inputs into results in civil, mechanical, electrical engineering, etc.

Hence, it becomes feasible to address Big Data Analytics practices under engineering concepts, since Analytics for Big Data presents characteristics that correspond to the classic elements of engineering, demanding mathematical and computational models to propose solutions to a variety of identified (open) problems.

Thus, in a conceptual approach, it is possible to consider Analytical Engineering a systematized way of using data, models, tools and people to transform problems related to data analysis for Big Data, developing the right solutions.

Analytical Engineering is determined as the application of technologies and techniques to a varied set of data through business, mathematical and computational models capable of delivering solutions (Analytics Deployments) proposed by an organization (Analytics Organization).

Reference [23] considers that Big Data Engineering includes advanced techniques that harness independent resources for building scalable data systems when the characteristics of the datasets require new architectures for efficient storage, manipulation, and analysis. This engineering concept for Big Data can be tied to Analytical Engineering practices.

In a conceptual approach, and possibly also from a practical view, Analytical Engineering as a transforming approach, basically of the data into knowledge, should be supported by data corresponding to the raw material, the basic input for this engineering proposal. Different components can be considered a part of Analytical Engineering to perform the proper transformation of data into knowledge. It is especially necessary to define models, and, in this case, models can be considered under business aspects, besides mathematical and computational aspects. Note that both have considerable depth, since the mathematical models to be used in these solutions are fundamentally based on probability theories and applied statistics. Computational models are also complex to fit to this approach and should be based on an extensive set of computing disciplines involving computing theory, applied computing, artificial intelligence, etc. [39]. It is also possible to verify that the models used by Analytical Engineering have a strong integrating aspect; for example, models that use Machine Learning theories integrate concepts of statistics under the computational domain characterized by statistical induction to generate predictive models [40]. The business models, however, have aspects not only linked to the domains of digital techniques and technologies, but to the fields of management sciences, business, economics, etc.

The Analytical Engineering approach also relies on the tools used for data transformation. Tools of various specificities arise constantly for data analysis in the Big Data domain, sometimes performing specific activities or requiring customizations or integrations with other tools to compose a better solution. However, a specific set of tools is needed for Analytical Engineering, either built in or acquired from suppliers; it is part of the assets of an organization that employs Analytical Engineering practices.

With the proper preparation, people are fundamental to this area as they are to other engineering areas. It is possible to consider the profile of an Analytical Engineer with the skill to develop models and tools and to transform data into knowledge. Characteristics similar to those of an Analytical Engineer can be verified in data scientists [41] and specific characteristics relate to the profile of a Chief Data Officer [42].

Analytical Engineering and its main components - the pillars of Analytical Engineering, are: Data, Models, Tools and People. Thus, the four pillars of Analytical Engineering should favor the implementation of a data-driven organization, which may be denominated an Analytics Organization. Possibly, the idealization of this organization should not be based only on these four pillars; however, in a conceptual way, such pillars should favor the institutionalization of an Analytics Organization.

The central aspect that can be considered for an Analytics Organization is its relationship with the organization's business areas, considering their links and integrations, relevant for materializing appropriate solutions. Considering Big Data Analytics, the institutionalization of an Analytics Organization is not feasible by traditional business models, based exclusively on the customer-vendor binomial. Integrated activities are constantly engaged and should be jointly developed [43], which favors the fact that the Analytical Engineer has to have multidisciplinary skills being able to work with statistical techniques, computational models associated with business model domains. Thus, an Analytics Organization is institutionalized aiming at delivering data-driven solutions to support the organization's business.

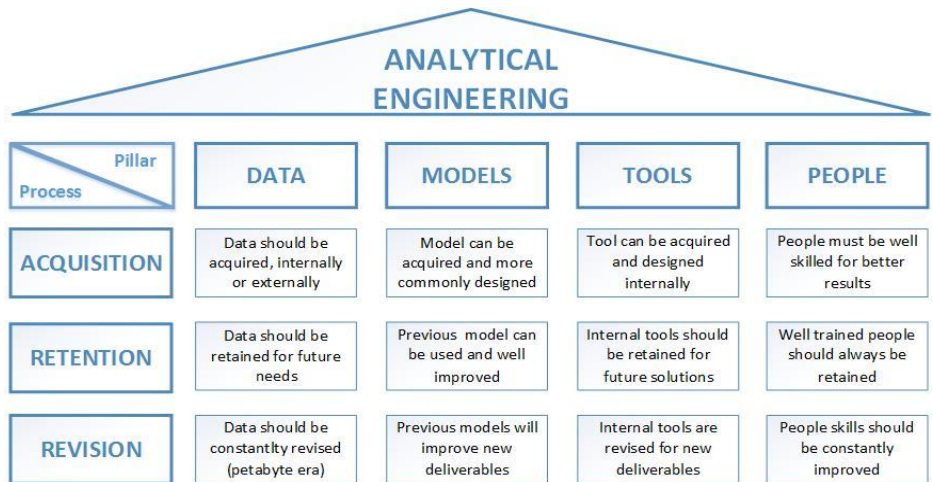

Fig. 1. Analytical engineering – Pillars and process groups.

Serving the business areas, an Analytics Organization must develop its deliveries, which can be characterized by Analytics Deployments. However, Analytics Deployments can be built under various constraints, such as: time, cost, technologies, data availability, resources, etc. Since the management of Analytics Deployments is one of the functions of an Analytics Organization and its group of engineers, management and operational structures in an Analytical Organization have unique aspects and need to be properly designed for each of them. Solutions that correspond to Analytics Deployments can be enabled as part of the operational activities of an Analytics Organization or can be further developed and delivered under the project principles. Several methodologies, such as the one presented by [44], are made available to Analytics; however, it is necessary for each Analytics Organization to define its managerial and operational processes to develop its Analytics Deployments.

Therefore, the activities and processes of Analytical Engineering are represented in the organization where they occur and through their results (Analytics Deployments). Thus, Analytical Engineering is able to delineate the management of an Analytics Organization, its processes and models, enabling the deployments of this organization.

Analytical Engineering focuses on its four pillars: data, models, tools and people; as well as its three process groups: Acquisition, Retention and Revision. Retention Process of data, models, tools and people favor the composition of an Analytics Organization. However, for these components to be retained, they must first be acquired (Acquisition Process), making it necessary to be constantly updated (Revision Process) to generate new deployments. The integration of these four pillars that define Analytical Engineering and its three process groups is presented in Fig.1.

As Analytical Engineering enables the institutionalization of an Analytics Organization and, consequently, the development of its Analytics Deployments, each of its pillars must undergo actions by the members of an Analytics Organization to propose the appropriate models of problems and solutions. Thus, the three main process groups of Analytical Engineering integrate the four pillars.

The definition of these process groups integrated to the four pillars represents the feasibility of Analytical Engineering for Big Data Analytics based on a conceptual approach, culminating in this Analytical Engineering proposal. Hence, each of these pillars that sustain Analytical Engineering is highlighted as follows, along with its process groups. Both, in an integrated way, allow managing an Analytics Organization for developing and maintaining its solutions that correspond to its Analytics Deployments.

**Data** correspond to the main asset of Analytical Engineering for Big Data. Data must be considered according to their properties, especially those that refer to volume, variety and velocity. These valuable inputs for this engineering proposal can be internally acquired or can be obtained externally, through various suppliers, such as government sectors, private companies, specific digital companies, etc., ensuring the condition of analysis and proper transformation to be used in business domains.

**Models** guarantee the transformation of data into knowledge. They are characterized by business models, mathematical and statistical models, computational models, and others, as necessary. They must be developed in an integrated way, considering multidisciplinary teams for better solutions modeling. The business model foresees actions to guarantee decision-makers favorable inputs from the various types of data, outlining the problem, to design a solution. Statistical and computational models can be presented in a more integrated way, especially since technologies and techniques, as well as specific tools, provide this type of integrated modeling [45]. These models often rely on concepts such as regression analysis, data correlation, Machine Learning algorithms, among others [46], [47] and [48], which enable the integrated definition of mathematical, statistical and computational models as presented by [49].

**Tools** become inherent in automated activities that provide solutions to data-driven problems. However, their scope requires specificity in the domains of technologies, that is, basic (or system) software or application software, both to enable the delivery of analytics [9]. Basic (or system) software mainly corresponds to the software technologies that favor the collecting, storing and managing data, considering technologies such as Apache Haddoop, various types of NoSQL DBMS, such as Big Table, Cassandra, Hbase [50], in-memory DBMS [8], and other technologies that determine data management and enable Analytical Engineering. In the specific domain of (big) data analysis, other technologies are presented: 1) for the analysis itself, and 2) for presenting results (data representation oriented tools) [51]. Examples of software applications or specific tools used emphatically for data analysis and visualization are Weka [52], R [53], Hazy [54] and MLBase [55] and [56], among others. The need to construct tools by an Analytics Organization, the acquisition of tools or even the definition of integrated or customized solutions [57] is part of a project in the Analytical Engineering domain that corresponds to the Analytics Projects.

**People** perform the actions, the processes that make deliveries (Analytics Deployments) of an Analytics Organization. Very specific skills have been designed for these tasks considering people with integrated domain in business, statistics and computing. These profiles with a multidisciplinary capacity favor faster response to these kinds of problems; however, their formation is based on a high degree of study as they need a high capacity of assimilation and mastery of multidisciplinary knowledge.

The four pillars of Analytical Engineering, briefly presented, should be treated under an extensive set of activities grouped into processes to provide the expectations of an Analytics Organization. These sets of activities using the four pillars of Analytical Engineering are part of the following process groups: Acquisition, Retention and Revision. For each of the pillars, the activities and procedures of Analytical Engineering can be part of these process groups, as detailed below.

The **Acquisition Process** corresponds to acquisitions that can be made by an Analytics Organization. Acquisitions occur in different ways for each of the pillars. Data acquisitions that correspond to the main input for this engineering proposal may occur internally or externally; acquisitions of models can be obtained from external sources, from the literature, from available research or can be completely designed internally; acquisitions of tools are supposed to be partially obtained externally; and, fundamentally, the 'acquisition of people' with the proper skill to work with these data, models and tools.

The **Retention Process** enables the assets of an Analytics Organization. Supposedly, under rigid configuration management, security, etc., the retention of data, if not all datasets, part of them, should enable future deployments. Models retention should occur because models may be useful for future solutions. Business models are likely to be almost completely altered, which interferes with the modifications of computational models, yet the retentions of previous models can collaborate with future solutions. Tools that are developed internally can provide new solutions, with few adjustments. In this case, tools correspond to the internally developed computational solutions to solve a given problem. 'People' are essential and must be 'retained' because they represent the intellectual capital that enables the management and operation of an Analytics Organization.

The **Revision Process** determines activities that may be performed by an Analytics Organization, and which are commonly performed by other engineering areas, which corresponds to reuse. This action requires data, models, and tools to be revised to propose new solutions; reusing these 'as-is' items may not allow for new solutions, but 'post-revision' adaptations may favor future deliveries. The 'revision of people' for Analytical Engineering is related to their profiles, which need to be constantly updated through courses, trainings, specific workshops, given the advancement of models, tools, technologies and techniques that are very common in this area.

The integration of components defined by Analytical Engineering allows delineating and maintaining an Analytics Organization. Under a conceptual approach, each of the components, the four pillars or specifically the three process groups proposed for Analytical Engineering, require specific details for each Analytics Organization. However, this level of detail is particular to each organization, which can be based on its internal models and structures, to institutionalize Analytics Organizations.

The feasibility of an Analytical Engineering proposal can be part of a Strategic Plan that can be based on a value chain structure. The value chain structures, the maturity models, and the other types of structures to support Big Data Analytics aforementioned are able to support a strategic view to determine the means of operating with Analytical Engineering.

Given that data analysis is a means for generating value in data-driven organizations, as aforementioned, the value that can be added by an Analytics Organization when implementing a structure based on the Analytical Engineering approach should be measured according to specific indicators and projects defined under Big Data Strategic Plan, and also based on the Analytics Deployments available. Also, indirectly, it is possible to verify the added value through decisions made on the basis of data-driven deliveries by Analytics Organizations.

V. CONCLUSION

Using a conceptual approach, qualifying what is necessary to implement and to institutionalize an organization focused on analytics practices for Big Data is what was sought with the research carried out and is what is presented by the proposal, which refers to Analytical Engineering for Big Data Analytics.

Analytical Engineering, possibly being treated as part of Big Data Engineering, refers to practices and processes that enable the implementation and institutionalization of an organization capable of handling the data analysis activities in the Big Data domain. That is, it favors the institutionalization of an Analytics Organization and the development of its solutions (Analytics Deployments).

Based on its four pillars (Data, Model, Tools, People) and its three process groups (Acquisition, Retention, Revision), Analytical Engineering determines the necessary components for developing solutions related to data analysis for the Big Data domain. These solutions (Analytics Deployments) delivered by an Analytics Organization should be promoted according to the activities grouped by the process groups defined by Analytical Engineering and

15

applied by an Analytics Organization.

The design and formatting of this conceptual proposal of Analytical Engineering was based on the principle that data analysis actions must provide value to an organization. Hence, value chain proposals were presented to relate value addition through analytics activities for Big Data. In a strategic vision, the value chain structure allows the organization to outline its steps to fit the Big Data era. More specifically and with greater operational dominance, Analytical Engineering provides four pillars and three process groups for institutionalizing analytics practices for Big Data in organizations expecting to become data-driven.

The Maturity Models and the other self-assessment structures that are presented favor the analysis of the capacity and maturity of organizations in handling data analysis for Big Data. These models are comprehensive and, sometimes, they do not detail the specific practices to be implemented, but are capable of directing the maturity of an Analytics Organization. However, when the maturity models highlighted herein are observed, they do not refer exclusively to analytics practices, but sometimes to the full cycle involving Big Data Systems.

In any case, the Analytical Engineering approach needs to be verified under practical aspects outlined by a guide, questionnaire, checklist, etc., to verify the practices carried out by the organizations to validate its four pillars in a practical view. This future work would determine an important result integrated to the proposal presented herein.

The deepening of each of the pillars and process groups that compose the conceptual approach of Analytical Engineering for Big Data Analytics as well a possible methodological view for analytical engineering can determine another relevant scope for future research regarding the engineering principles applied to Big Data Analytics.

A specific research that considers the relationship between Project Management principles and Analytics to favor the management of Analytics Projects capable to determine the development of Analytics Deployments, by multidisciplinary teams, may reveal relevant future work to determine or to improve the Analytics Management approach.

ACKNOWLEDGMENT

We thank Programa de Educação Continuada em Engenharia (PECE) (Engineering Continuing Educational Program) of Polytechnic School of University of São Paulo for supporting this work.

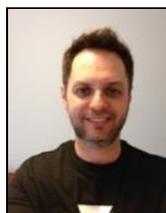

**Rogério Rossi** holds B. Sc. in mathematics by the University Center Foundation Santo André (1992) and he also holds a M.Sc. (1998) and Ph.D. (2013) in electrical engineering (computer engineering), both by Mackenzie Presbyterian University.

He is a visitor professor for Engineering of Polytechnic School at University of São Paulo for MBA Programs (Software Technology and Internet of Things). He has realized research on the fields of software quality, business in technology, and information technology for education. He has some published papers in these areas. Nowadays, his research interests include Analytics for Big Data and Data Science.

Dr. Rossi is an IACSIT member and he worked as a reviewer for some international conferences. He is a reviewer of ISI Journal (IJELL) and he also presented his works in the ISI Conferences in Montreal, Canada (2012) and Porto, Portugal (2013); and ICCSIT'2019 in Barcelona, Spain.

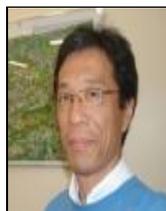

**Kechi Hirama** is an associate professor of Polytechnic School of University of São Paulo, Brazil, in Department of Computer Engineering. He holds the B.Sc., M.Sc., Ph.D. and associate professor degrees in Computer Engineering from Polytechnic School of the University of São Paulo. He worked in the control systems and industry automation areas in research organizations and currently is responsible for Complex System Group in Department of Computer Engineering. His interests include System Dynamics, Complex Networks, Big Data and Internet of Things.